\documentclass[modern, intlimits]{aastex631}

\usepackage[utf8]{inputenc}

\newcommand\latex{La\TeX}
\usepackage{xcolor}
\usepackage{natbib}
\usepackage{graphicx}
\usepackage{verbatim}
\graphicspath{ {./images/} }
\usepackage{amsmath} 
\usepackage{mathtools}
\newcommand{\be}{\begin{equation}}
\newcommand{\ee}{\end{equation}}
\usepackage{mathrsfs}
\usepackage{algorithm,algorithmic}
\usepackage{enumitem}

\begin{document}
\title{SURF Report: High Accuracy Methods for Computing Gravitational Potential and Gravitational Force Fields Near the Surface of Irregularly Shaped 3-Dimensional Bodies\footnote{Copyright: © 2024. California Institute of Technology}}
\author{\textbf{Author: }Thomas MacLean}
\affiliation{California Institute of Technology, Pasadena, CA, 91125, \textcolor{blue}{\underline{\textbf{tmaclean@caltech.edu}}}}
\author{\textbf{Mentor: }Prof. Alan H. Barr}
\affiliation{California Institute of Technology Department of Computing and Mathematical Sciences, Pasadena, CA, 91125}

\section{Abstract}
Accurate gravity field calculations are necessary for landing on planets, moons, asteroids, Earth's ``minimoons," or other irregularly shaped bodies, but current methods become increasingly inaccurate and slow near the surface. We present high accuracy, fast methods for computing gravitational potential and gravitational force fields, which are needed for future space missions. Notably, gravitational force and potential computations are simplified, with high accuracy enhanced by bringing the derivative inside the gravitational potential integral. In addition, we present a new gravitational field “calculus,” which lets us combine simpler potentials and force fields to create more complex ones without accuracy loss. Several examples are provided, for instance, where we subtract different shapes from a spherical body making a variety of craters. The calculus will also work well with volumetric octree methods. Additionally, we use new bounds in the gravitational potential integral, to avoid trying to fit smooth basis functions to non-smooth curves, and harness new computational tools where tasks can be migrated to GPUs. We also have found that cylindrical coordinates can have special advantages in tailoring shape models. We have created a series of algorithms and preliminary MATLAB and Mathematica toolboxes utilizing these methods and the gravitational calculus. These methods are newly customizable for necessary high-accuracy gravity computations in future missions planned by JPL and other space agencies to navigate near irregularly shaped bodies in the solar system.

\section{Introduction}
Computing the gravitational potential field $V$ and gravitational force field $F$ are highly applicable and significant results for navigational mathematics, physics and engineering necessary for landing and modeling near-body dynamics. Throughout this paper we refer to ($V, F$) as the ``gravity field." While the reader may assume that fast, accurate computations of the gravity field near non-spherical objects are already available, this is not the case. Surprisingly, current faster state-of-the-art gravitational force computations exhibit orders of tens of percent error near asteroid and moon surfaces \citep{scheeres2016orbital}, \citep{SCHEERES2019}, \citep{2014Takahashi} and are insufficient for accurately representing the dynamics. Most of these surface gravity models create errors on the order of tens of percent for bodies that stray from spherical symmetry, and the more aspherical the shape is, the more inaccurate the gravitational force models near the surface are \citep{2014Takahashi}. We foresee that our research opens the door to developing strategies for faster and far more accurate navigation near the surface of bodies.

In particular, space mission design and flight design are increasingly in need of fast accurate surface gravity force computations for the sake of landing, as well as for general navigation near the surface of real bodies. As landing on or flying near asteroids, moons, and other non-spherical bodies in space becomes increasingly pertinent, our current improvement of accuracy in gravity computations near the surface of realistic, non-spherical bodies will be increasingly beneficial. 

\begin{figure}[h!]
\centering
\includegraphics[scale = 1]{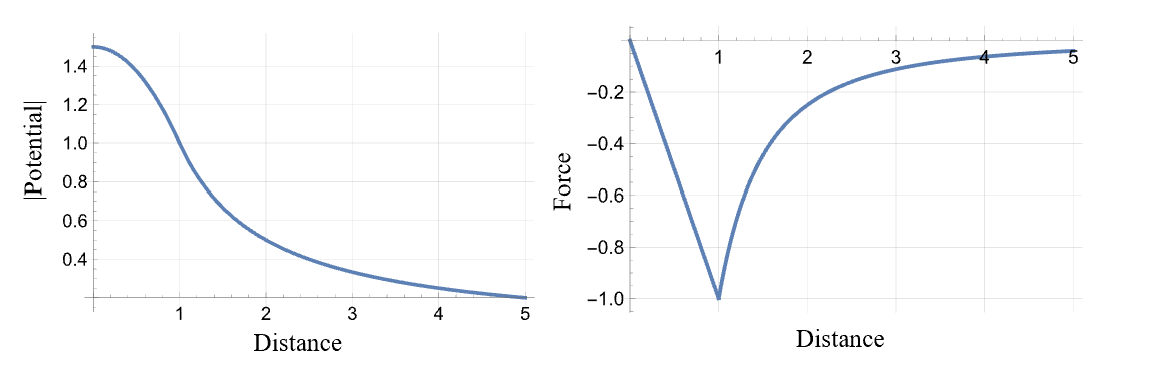}
 \caption{Gravity field for uniform density radius R=1 sphere along a line from the sphere center, in any direction, in non-dimensional units. Gravitational potential $V_S$ is $C^1$ continuous but not $C^2$ continuous. Note the cusp in the force function $F_S$, which makes $F_S$ unsuitable for representation with smooth basis functions. The x-axis is the distance from the center of the sphere in units of sphere radii. The left plot is the closed-form absolute value of the gravitational potential of the sphere, $|V_S|$, along any two-dimensional line from the center of the sphere ($V_S$ shown in Equation \ref{eq:spherep}), and the right plot is the closed-form gravitational force of the sphere, $F_S$, along any two-dimensional line from the center of the sphere (shown in Equation \ref{eq:spheref}).}
    
    \label{fig:sgrav}
\end{figure}

\textbf{New Methods: }We present new methods for fast, reliable navigation near realistic irregularly shaped bodies that can be used for determining the optimal navigation strategy in a given mission. We present a number of methods that can be used for highly accurate, fast and reliable calculations, with special focus on our ``calculus" of gravitational potentials and forces. Additionally, our methodology avoids issues in the ``conventional methods" by taking the derivative of the potential integrand, solving the definition of the gravitational force directly, and not working with smooth basis functions to fit to non-smooth curves. See Figure \ref{fig:sgrav} for a non-smooth force curve that is not suitably fitted by smooth basis functions. This work has been accepted as California Institute of Technology Jet Propulsion Laboratory New Technology Report 53321 \citep{JPLNTR53321}.

We address a current issue that lies in computing sufficiently accurate derivatives of standard potential functions to obtain accurate gravitational force fields near the surface \citep{Werner2010traj}, \citep{scheeres2016orbital}. In the example case of the Mars moon Phobos, computational methods to derive the gravity field with accuracy and speed are crucial to allow for reliable navigation, and the gravitational force calculations on Phobos are especially relevant where the atmosphere is thin and asymmetries make the body highly non-spherical \citep{WILLNER2010541}, \citep{SCHEERES2019}. 

At the core of our new methods, we focus on matching solutions closely to the ``true" gravitational potential field $V$ and gravitational force field $F$. Letting $\nabla$ denote the gradient, the ``true gravity field definitions," with no additional assumptions, are given by Equation \ref{eq:defpot} and Equation \ref{eq:defforce} for gravitational potential $V$ and gravitational force $F$, respectively, evaluated at any point $\overrightarrow{\mathbf{x}} \in \mathbf{R}^3$.

In Appendix B, Section \ref{sec:ap2}, we present insights into the challenges that arise in the conventional methods, and we also present an implementable algorithm (B.1) derived from \cite{2014Takahashi} which gives the previous state-of-the-art fast computation method. 

\begin{align} \label{eq:defpot}
            V(\overrightarrow{\mathbf{x}}) &=  \int_{\mathbf{R}^3} \frac{G\rho(\overrightarrow{\mathbf{r}})d\mathbf{v}(\overrightarrow{\mathbf{r}})}{|\overrightarrow{\mathbf{x}} -\overrightarrow{\mathbf{r}}|} \\ \label{eq:defforce}
            F(\overrightarrow{\mathbf{x}}) &=  -\nabla V(\overrightarrow{\mathbf{x}})
        \end{align}

\section{Methods} \label{sec:methods}
Looking directly at the definition of the gravitational potential $V$ in real space induced by a shape $\mathbf{S}$ with density function $\rho(\overrightarrow{\mathbf{r}})$, given by Equation \ref{eq:defpot}, we can compute the gravitational potential field $V$ induced by the shape $\mathbf{S}$. Integrating the integrand $ \frac{G\rho(\overrightarrow{\mathbf{r}})d\mathbf{v}(\overrightarrow{\mathbf{r}})}{|\overrightarrow{\mathbf{x}}-\overrightarrow{\mathbf{r}}|}$ over $\overrightarrow{\mathbf{r}} \in \mathbf{R}^3$, where $G$ is the gravitational constant, $\rho(\overrightarrow{\mathbf{r}})$ is the density function which gives the density of the body at a given point in space, $d\mathbf{v}(\overrightarrow{\mathbf{r}})$ is the volume element, and $|\overrightarrow{\mathbf{x}}-\overrightarrow{\mathbf{r}}|$ is the distance between input vector $\overrightarrow{\mathbf{x}} \in \mathbf{R}^3$ and $\overrightarrow{\mathbf{r}}  \in \mathbf{R}^3$, we derive the expression for the true gravitational potential field $V$ induced by the shape $\mathbf{S}$.  

The integration process just described, where we integrate as in Equation \ref{eq:defpot} over the shape $\mathbf{S}$, is the most direct way of computing the gravitational potential field, and is derived directly from the gravitational potential definition. Staying ``near" the direct definition presents a powerful way of limiting error in our solution; we can derive closed-form expressions for the gravity field for certain shapes $\mathbf{S}$ by conventional integration tricks. 

The gravitational potential $V_S$ for a uniform density sphere is a continuous $C^1$ function (see Figure \ref{fig:sgrav}), but its derivative exhibits a fundamental issue near the surface. This restricts the standard methods from accurately deriving the gravitational force $F_S$ for the sphere, which is the derivative of the potential. The graph of gravitational force may seem to be smooth despite its underlying behavior as a $C^1$, not $C^2$, function, as demonstrated by Figure \ref{fig:sgrav}. 

Closed-form expressions for the uniform density sphere's gravity field can be derived by substituting the sphere into Equation \ref{eq:defpot}, where over $\mathbf{R}^3$ the density $\rho({\overrightarrow{\mathbf{r}}})$ is 0 outside of the sphere and 1 inside the sphere.

\begin{equation} \label{eq:spherep}
V_{S}(\overrightarrow{\mathbf{x}}) = \begin{cases}
-\frac{GM}{||\overrightarrow{\mathbf{x}}||}, & \text{if } ||\overrightarrow{\mathbf{x}}|| > R \\
-\frac{GM}{2R} \left( 3 - \frac{||\overrightarrow{\mathbf{x}}||^2}{R^2} \right), & \text{if } ||\overrightarrow{\mathbf{x}}|| \leq R
\end{cases}
\end{equation}

\begin{equation} \label{eq:spheref}
F_{S}(\overrightarrow{\mathbf{x}}) = \begin{cases}
-\frac{GM}{||\overrightarrow{\mathbf{x}}||^3} \overrightarrow{\mathbf{x}}, &  ||\overrightarrow{\mathbf{x}}|| > R \\
-\frac{GM}{R^3} \overrightarrow{\mathbf{x}}, & ||\overrightarrow{\mathbf{x}}|| \leq R
\end{cases}
\end{equation}

See Equations \ref{eq:spherep} and \ref{eq:spheref} for closed-form gravity field expressions about the uniform mass sphere with $F_S$ denoting the gravitational force field about the sphere and $V_S$ denoting the gravitational potential field, where we take the positive direction to be pointing inwards and let $||\overrightarrow{\mathbf{x}}||$ denote the norm of vector $\overrightarrow{\mathbf{x}} \in \mathbf{R}^3$.

See Figure \ref{fig:sgrav} for a plot of these expressions in any given direction from the center of the sphere, where the gravitational force has a cusp at the sphere boundary.

\textbf{New Methods to Eliminate Conventional Errors:} Real shapes of interest typically have other characteristics beyond spherical symmetry and constant density that may make dynamics near the surface additionally complex. We present a method whereby integrating Equation \ref{eq:defpot} directly, there is no computational noise which may come from unsuitable basis functions, and taking the gradient of Equation \ref{eq:defpot} prior to integration results in a highly accurate numerical or closed-form gravitational force field. See Algorithm A.1 in Section \ref{sec:ap1} for a new procedure for gravity field computation via integration which stays near the true gravity field definitions given by Equations \ref{eq:defpot} and \ref{eq:defforce}.

\textbf{Symbolic and Numeric Integration: }The triple integrals for computing gravitational potential and force, as given by Equations \ref{eq:defpot} and \ref{eq:defforce} evaluated for a given shape $\mathbf{S}$, at a point $ \overrightarrow{\mathbf{x}} = (x,y,z)$, may have one, two, or three symbolic integrations depending on the choice of coordinate system $(\phi_{1}(x),\phi_{2}(y),\phi_{3}(z))$, where choosing a convenient coordinate system can significantly speed up the computation while retaining accuracy. Each symbolic integration speeds up the computation because it provides a closed-form expression. This choice of coordinate system is motivated by selection of coordinates which exploit symmetries. We select the coordinate system $(\phi_{1}(x),\phi_{2}(y),\phi_{3}(z))$ which allows for a higher number of symbolic integrations in order to maintain speed in the computation and limit numerical integration. 

\begin{figure}
\centering
\includegraphics[scale = .625]{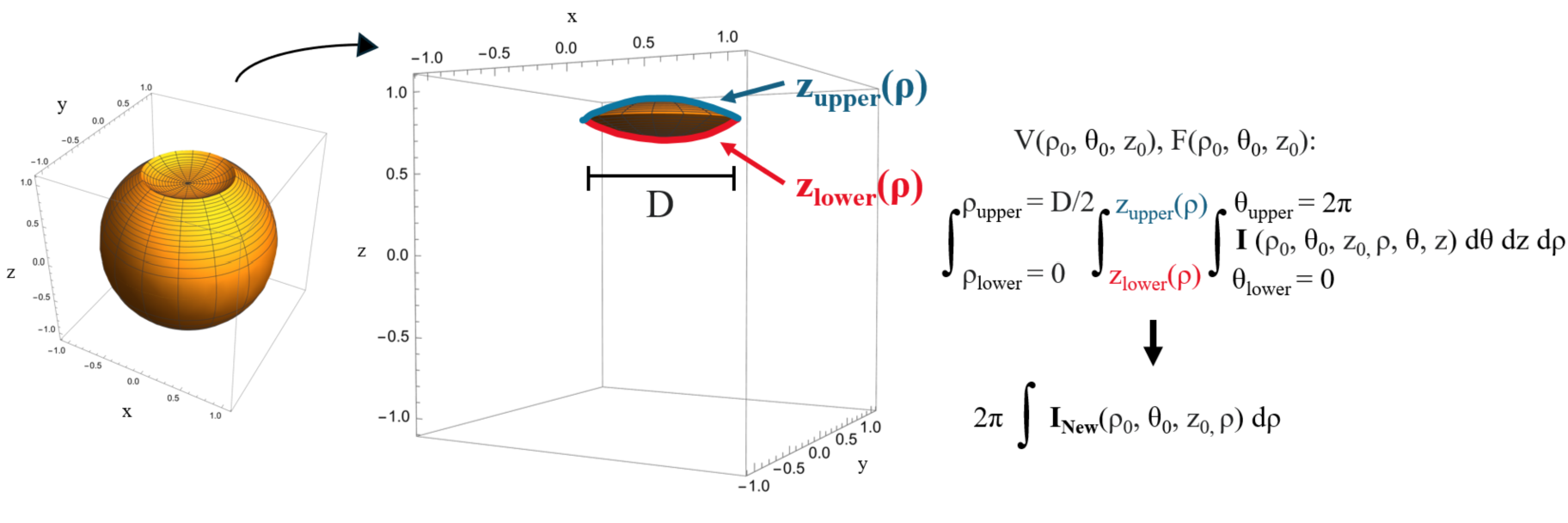}
 \caption{Computing the gravity field of bowl-shaped ``simple" craters can be done with convenient bounds in cylindrical coordinates. The final integral involves only one numeric integration (here, over $\mathbf{I_{New}}$ in $ \rho$ as opposed to integrating the original integrand $\mathbf{I}$ over $\rho, \theta, z$) and two symbolic integrations, due to the symmetry in $\theta$ and ability to write bounds of $z$ as a continuous function of cylindrical coordinate $\rho$. See \cite{2003K}, \cite{2007craters} for other crater classifications, which can also be modelled conveniently in cylindrical coordinates, due to similar properties of symmetry in $\theta$ and ability to write bounds of $z$ as a continuous function of cylindrical coordinate $\rho$.}
    
    \label{fig:bounds}
\end{figure}

We find that in cases where there is symmetry about the z-axis and the height can be written as a function of the radial distance from the center, the use of conventional cylindrical coordinates $(\rho, \theta, z)$—where $\rho$ is radial distance in the xy-plane from the origin, $\theta$ is the angle relative to the x-axis in the xy-plane, and $z$ is height—is convenient. This choice of coordinates lends way to two symbolic integrations, following the definitions in Equations \ref{eq:defpot} and \ref{eq:defforce}. Spherical symmetry leads to a constant factor from integration in $\theta$, and we can write the bounds in $z$ as a function of planar coordinate $\rho$, which allows for symbolic integration in $z$. These properties hold for many notable volumes which may represent true surface features, such as craters. An example of the setup for computing the gravity field of a crater volume, where integration is simplified with two symbolic integrations via choice of cylindrical coordinates, is shown in Figure \ref{fig:bounds}.

\begin{figure}[h!]
\centering
\includegraphics[scale = .6]{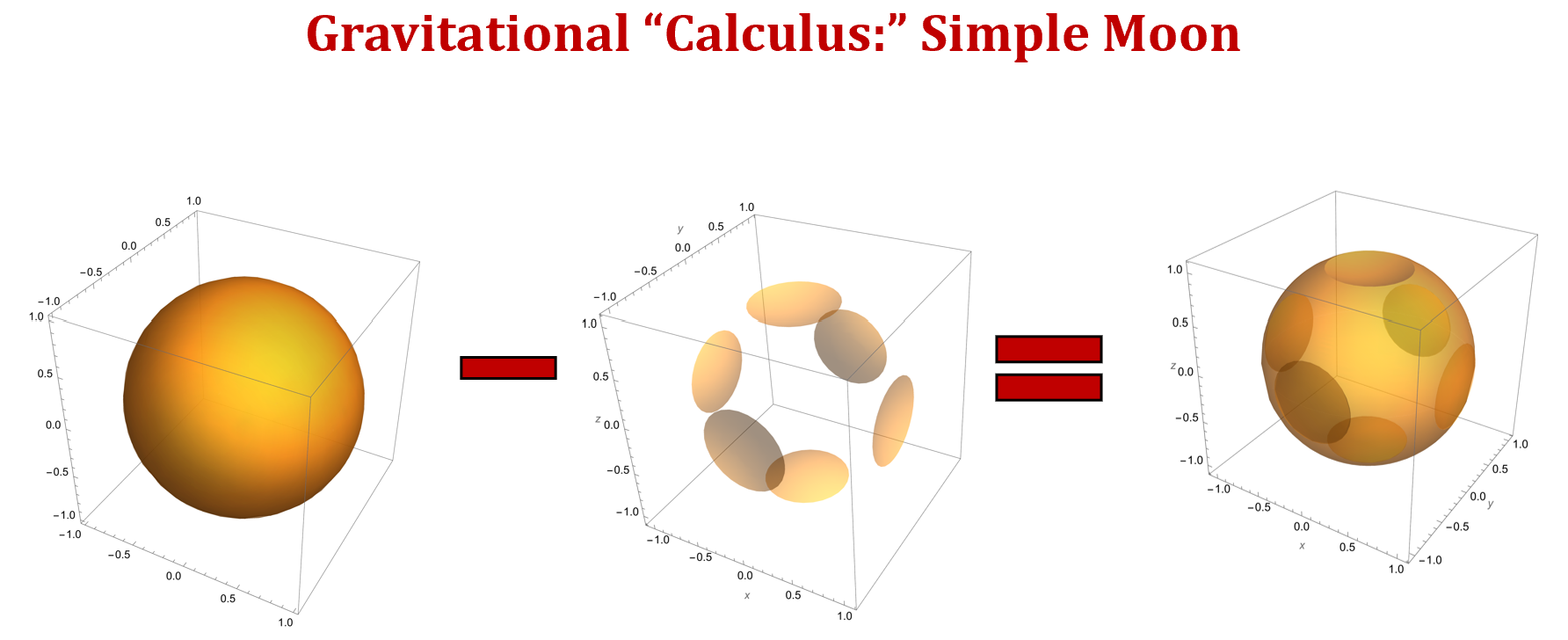}
 \caption{\textbf{Example of Gravitational Calculus:} Sphere with several bowl-shaped ``simple" craters, which models a simple moon. The gravity field of the final shape is equal to the gravity field of the sphere minus the gravity field of the ``simple craters."}
    
    \label{fig:moon}
\end{figure}

\begin{figure}[h!]
\centering
\includegraphics[scale = .6]{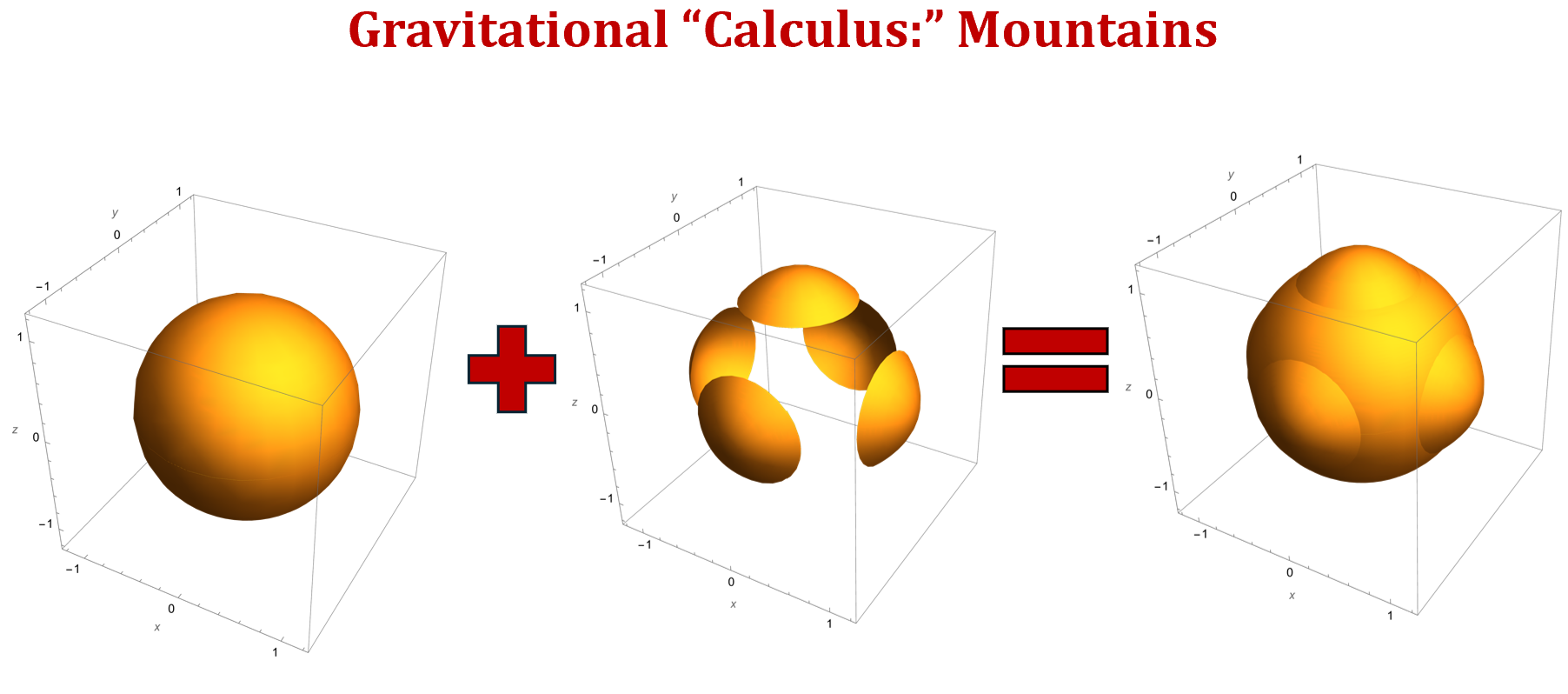}
 \caption{\textbf{Example of Gravitational Calculus:} Sphere with mountain appendages. The gravity field of the final shape is equal to the gravity field of the sphere plus the gravity field of the mountain volumes. The gravity field of the mountain volume need only be computed once since each mountain is a rotation of the others.}
 \label{fig:spikyball}
\end{figure}

\begin{figure}[h!]
\centering
\includegraphics[scale = .6]{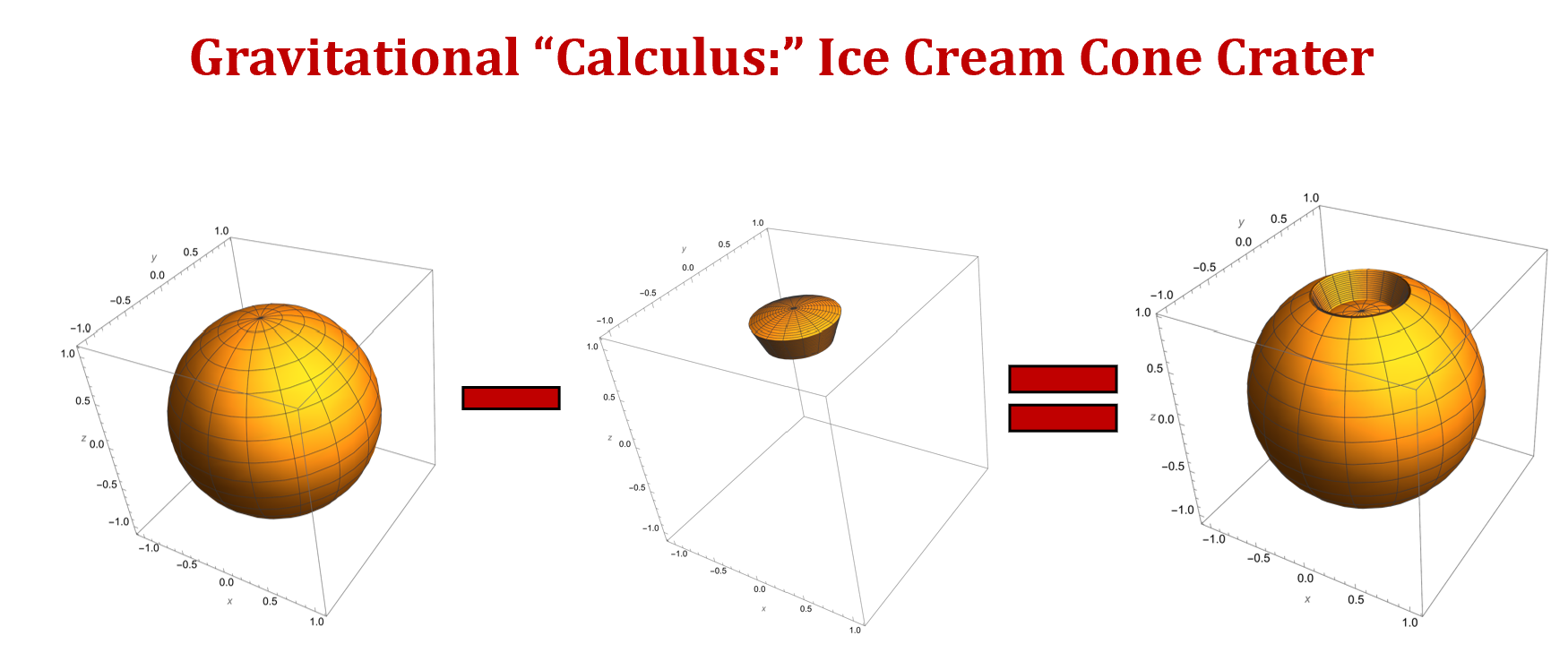}
 \caption{\textbf{Example of Gravitational Calculus:} Sphere of radius R with an ``ice cream cone crater." The gravity field of the final shape is equal to the gravity field of the sphere minus the gravity field of the crater volume. The spherical coordinate bounds of the crater volume, the piece subtracted from the sphere, are $\frac{3R}{4}$ to R in radial distance coordinate, 0 to 2$\pi$ in $\theta$ coordinate, and 0 to $\frac{\pi}{8}$ in $\phi$ coordinate.}
    \label{fig:icecream}
\end{figure}

\textbf{New ``Gravitational Calculus:" }We present uses of our new ``gravitational calculus," where computation of disjoint shapes' gravity field can be translated or rotated, then added or subtracted to the overall gravity field. See Figures \ref{fig:moon}, \ref{fig:spikyball}, and \ref{fig:icecream}, for examples of shapes with craters and mountains. The operations of translation and rotation via the gravitational calculus are given by Algorithm A.3.1 and A.3.2 in Appendix A, Section \ref{sec:ap1}. 

Certain closed-form solutions can be used in our new type of ``gravitational calculus." Closed-form solutions are possible when a convenient choice of coordinate system allows for three symbolic integrations. This is the case for the gravity field about the sphere. Additionally, highly accurate numerical computations can be used as a part of the calculus, where once a shape's gravity field has been computed, use of the calculus allows for translation or rotation of the shape. In cases where one or more numerical integrations are computed to represent the gravity field of a shape $\mathbf{S}$, although not exact, that gravity field can also be utilized further via the gravitational calculus with rotation or translation. 

\begin{figure}[h!]
\centering
\includegraphics[scale = .5]{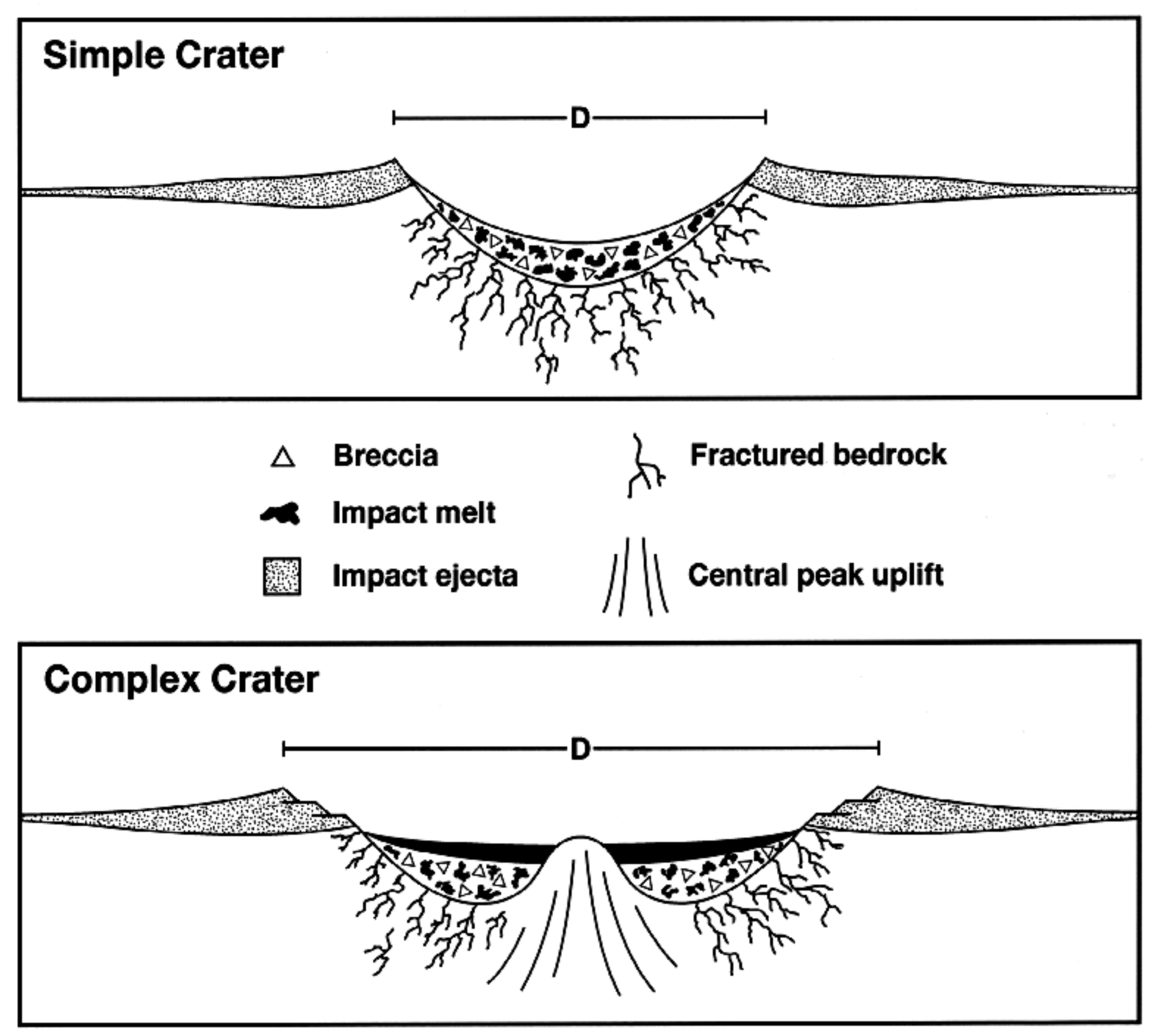}
 \caption{Scientific ``simple crater" and ``complex crater" classifications, figure from \cite{2003K}. Such craters may exist on the shape of interest. The crater's contribution to the gravity field can be realized via choice of coordinate system, new gravitational potential integral bounds, and use of the gravitational calculus.}
    
    \label{fig:sc}
\end{figure}

\textbf{Navigation: }For navigation purposes, it is often necessary to investigate crater-filled shapes. Scientific classifications exist for craters, with an example of a simple and complex crater shown in Figure \ref{fig:sc}, while other craters may have even more complicated shapes, which carry other classifications \citep{2003K}. Impact craters may have specific morphological shapes including mounds, flat floors, or concentric shapes \citep{2007craters}.

A potential shape to investigate, in order to model a given crater, would have some bounds in radius, within a choice coordinate system (for instance, spherical coordinates for the following consideration), that are functions $c_{lower}(\phi)$ and $c_{upper}(\phi)$ of $\phi$ chosen in such a way of closely representing the ``real crater."  This can be used to model real craters that form and lie in real bodies such as Phobos. See Figure \ref{fig:icecream} for a sphere with a ``test crater" which algebraically simplifies since $c_{lower}(\phi)$ and $c_{upper}(\phi)$ are constants, where use of the gravitational calculus presents a gravity field computation for an asteroid-like shape.

We present additional convenient applications of the gravitational calculus in Figures \ref{fig:moon} and \ref{fig:spikyball}, where a shape is rotated for use in computing the final shape's desired gravity field, and the gravitational calculus makes the computations convenient. The gravitational calculus allows us to model mountains, craters, or other surface features. 

\textbf{Octree Methods:} We present another new method, which utilizes octrees for shape and density representations to compute the gravity field in Algorithm A.2 of Appendix A, Section \ref{sec:ap1} \citep{MEAGHER1982}. Algorithm A.2 provides a guide for implementing octree representation. The use of quad tree algorithms has been examined in aerospace applications for shape representation of satellite mapping coverage \citep{lohockney2003}, and many of the same benefits coming from this form of shape representation of completeness and efficiency may apply in our case of gravity field computations, given our gravitational calculus. The octree is a three-dimensional generalization of the quad tree representation, and this method allows for complete shape representation in three dimensions.

\medskip 

\section{Results} \label{sec:results}
We have demonstrated that the nature of the gravitational force, with non-smooth behavior, encourages use of new fast methods for high accuracy, yet efficient, computations, which we propose. These methods are summarized by the algorithms that are stated in the following section. 

\begin{figure}[h!]
\centering
\includegraphics[scale = .84]{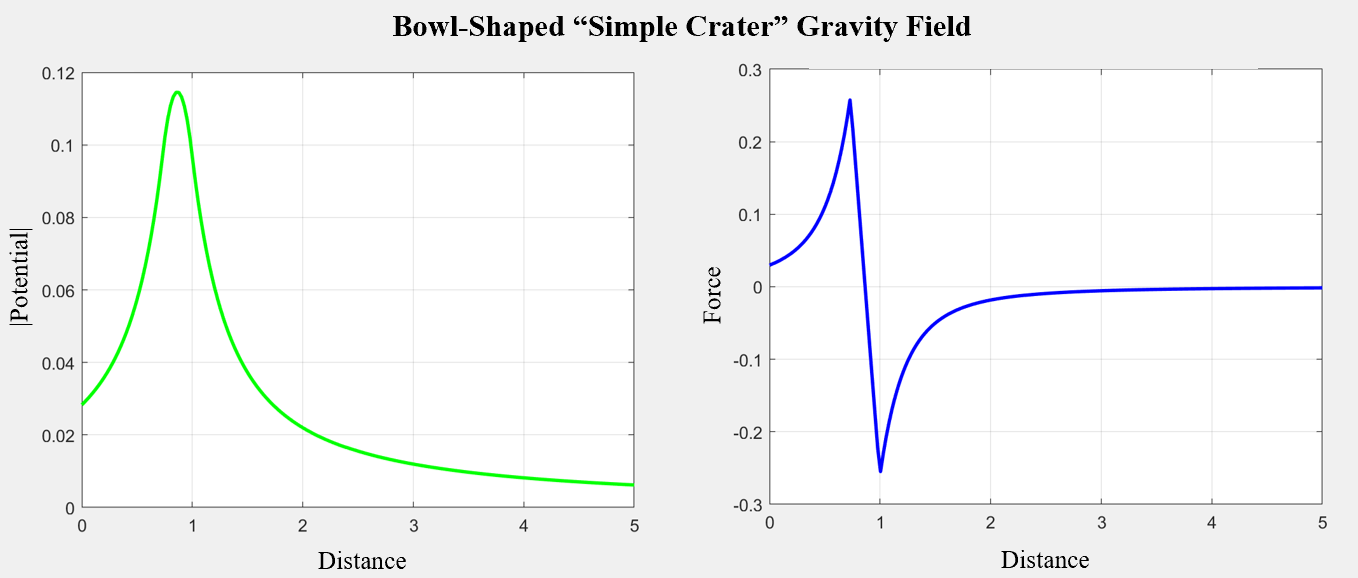}
 \caption{Absolute value of gravitational potential ($|V|$) versus distance, and gravitational force ($F$) versus distance of bowl-shaped ``simple" craters in direction of North Pole. Figure made originally in Mathematica and reproduced in MATLAB using our new GravToolbox Package, where use of cylindrical coordinates gives a convenient computation.}
    
    \label{fig:bowlgf2}
\end{figure}

While it has been previously demonstrated that within the minimum Brillouin sphere, where all mass is enclosed, the force function is not smooth by nature \citep{costin2020}, \citep{SPRLAK2021103739}, we find that, in fact, anytime a body's section crosses a mass boundary, such as in craters or mountains, there can be cusps arising in the true force function. See Figure \ref{fig:bowlgf2} for an example of the gravity field about a crater modeled to fit the form of a ``simple crater," where we have a seemingly innocent potential function whose derivative has two cusps, which correspond to the points on the boundary of the crater volume. 

We have developed tools for integrating over certain ``crater sections" by directly integrating the potential definition in Mathematica and MATLAB code. See Appendix (Section \ref{sec:ap1}).

\textbf{Algorithms: } We have constructed a set of algorithms which guide computation of gravitational potential field $V$ and gravitational force field $F$ for a given shape $\mathbf{S}$. See Appendix A, Section \ref{sec:ap1}, for full set of new method algorithms. Algorithm A.1 states a workflow for computing the gravitational potential and force fields. The algorithms that follow, in Section \ref{sec:ap1}, present direct methods involving integration in solving for $V$, $F$, the octree method (Algorithm A.2), gravitational calculus operator properties (Algorithms A.3.1, A.3.2), and tools for simplifying computations using the gravitational calculus and properties of the gravitational potential (Algorithms A.4, A.5). See Section \ref{sec:applications} for additional future pathways for computing gravity fields. In Appendix B, we condense the previous conventional methods into a convenient algorithm; Algorithm B.1 re-states in brief, implementable fashion, a state-of-the-art method as presented in \cite{2014Takahashi}. This algorithm is also implemented in Mathematica notebook ``Potential Reconstruction Algorithm," created for demonstration of the previous methods versus the new methods. Figure \ref{fig:error} demonstrates the success of following the new proposed gravity field workflow in the case of the uniform mass sphere, where there is only numerical error, whereas in the conventional methods, there is additional error (See Appendix B, Section \ref{sec:ap2}).

\textbf{MATLAB Tools: } We have a preliminary MATLAB package, GravToolbox, which computes, and plots in a 2-dimensional cross-section, the numerical potential of a specified shape. Inputting the bounds which represent the crater shape (as a spherical section which can be expressed in terms of numerical bounds for $(r, \theta, \phi)$ in spherical coordinates) into our functions, we can compute a (highly accurate) numerical cross-section of the potential and force field. Figure \ref{fig:bowlgf2} shows the gravity field computation for a uniform density crater modelled to fit the shape of a ``simple crater."

\textbf{Mathematica Tools: } We have preliminary Mathematica notebooks which compute and plot in a 2-dimensional cross-section, the numerical potential of a specified shape, as done in the MATLAB package GravToolbox. We also have an example notebook ``Potential Reconstruction Algorithm" which demonstrates Algorithm 1 and error measurements (see Appendix, Section \ref{sec:ap1}).

\section{Applications \& Future Work} \label{sec:applications}
As we have begun to build the calculus for the gravitational potential and gravitational forces, we have found that many of the potential integrations become algebraically complicated, leading to long run times and challenging computations.  Taking the derivative of the potential's integrand prior to integration to find the gravitational force field can lead to a more challenging integral than that of the potential integral. However, these run times can be sped up by some optimal combination of pre-computed octrees or pre-computed ``definition integrals" in convenient coordinate systems and bounds for shapes, each of which need only be computed once. Our methods, at their finest, are highly accurate, but potentially slow in such a way that certain computations may be infeasibly slow without GPUs. This is the issue with the polyhedron model \citep{WERNER19971071}, \citep{Werner2010traj}. However, we anticipate that computation of shapes in parallel using GPUs creates a modern, sufficiently fast method which retains high accuracy up to specified tolerance. We can ensure a desired accuracy, up to some tolerance, in the case of computing numerically by computing the exact potential at a sufficiently large number of points and interpolating. Similarly, with use of octrees, we can continue the algorithm up to construction of a shape that is sufficiently close to the true shape of interest. We also anticipate that there may be special cases where a scaling property could be used with the gravitational calculus to speed up computations. 

In the example case of Phobos, a highly accurate form, we predict, would require adaptability to a wide span of shapes that constitute the body in union, and this may involve numerical and octree computations in conjunction with our potential ``calculus" (see \citep{WERNER19971071} for Polyhedron approximation of non-spherical bodies). The approach would also extend to navigation near the Earth's ``minimoons." These accurate computations model gravitational force fields which are very ``cusp-like" in behavior, and are not limited to smooth functions. 

Further work involves a deep investigation into how changing integration order, reformatting integrands, optimizing code, finding shapes that represent ``true craters" in ways that are more conveniently derivable, changing coordinate systems, or using more powerful computing can be combined in such a way that creates a directly implementable algorithm for space agencies to use. Additionally, new algorithms must be able to model non-constant density distributions, which may be done using the octree method in conjunction with machine learning algorithms, to fit a power series density function to each closed-form box in the octree. Our GravToolbox package can be expanded into a re-usable set of code for navigational purposes and our algorithms in Section \ref{sec:ap1} can also be implemented for specific gravity field computations in missions requiring high accuracy navigation near bodies.

\textbf{Additional Methods:}
It may be possible to use a different solution to Poisson's Equation $\nabla^2U = -4\pi G \rho$ using complete basis functions such as wavelets \citep{2023Wave} and a more complete representation of the problem itself as the single differential equation given by Equation \ref{eq:completes} where for an ellipsoid with axes a, b, c, in the x, y, and z directions respectively, $f(x,y,z) = 1-\frac{x^2}{a^2}-\frac{y^2}{b^2}-\frac{z^2}{c^2}$. 

\begin{equation} \label{eq:completes}
\nabla^2U = H(f)
\end{equation}

One solution method in the new formulation involving the Heaviside function representation could be to use a Monte Carlo algorithm which samples points randomly. Other investigations could involve a Fourier Transform Method. Others could investigate other distributions to represent the solution. Speed-up and parallel computations may utilize GPUs. 

\section{Acknowledgements} \label{sec:ack}
Thomas MacLean is supported as a William H. and Helen Lang Summer Undergraduate Research Fellow with funding support provided by William H. Lang and Helen Lang and the California Institute of Technology (Caltech) Student-Faculty Programs Office. Dr. Martin Lo at the NASA Jet Propulsion Laboratory has been a vital source of ideas and collaborator for the research. This research is directly funded and supported by the Caltech Barr Graphics Lab. The NASA Jet Propulsion Lab has provided funding, allocated via the Mission Design Section and made possible by Dr. Jon Sims and Dr. Martin Lo. The research was carried out in part at the Jet Propulsion Laboratory, California Institute of Technology, under a contract with the National Aeronautics and Space Administration (80NM0018D0004).

© 2024. California Institute of Technology. Government sponsorship acknowledged.

\vspace{5mm}

\software{\latex, Mathematica, MATLAB}
\clearpage
\section{Appendix A} \label{sec:ap1}
\noindent\fbox{%
  \parbox{\dimexpr\linewidth-2\fboxsep-2\fboxrule\relax}{ \textbf{\large{Algorithm A.1: New Proposed Workflow for Gravitational}} \\ \textbf{\large{Potential $V$, Gravitational Force $F$ Computation}}\label{tab:alg1}
  \newline ||||||||||||||||||||||||||||||||||||
  \begin{enumerate}[leftmargin=*]
    \item   \textbf{Prepare integral definition of gravitational potential in closed-form.} Represent potential $V(\overrightarrow{\mathbf{x}})$ as the integral definition, see Equation \ref{eq:defpot}.
    \item  \textbf{Partition and choose coordinate systems.} Partition the shape $\mathbf{S}$ into sections consisting of base shape $\mathbf{B}$, other appendage or intersecting shapes (see A.5 for use of intersections) \{$\mathbf{S}_i$\}$_i$ where computations will be done. We select coordinate systems $\{(\phi_{1i}(x),\phi_{2i}(y),\phi_{3i}(z))\}_i$ to simplify integrations involved. Our goal is to optimize over low runtime, error, and numerical computations. See Section \ref{sec:methods} for discussion about convenience of selection of coordinates given density function $\rho(\overrightarrow{\mathbf{r}})$ of shape $\mathbf{S}$.
    \item \textbf{Use Gravitational Calculus.}  In computing gravity fields due to base shape $\mathbf{B}$ and appendages \{$\mathbf{S}_i$\}$_i$, if suitable, refer directly to Algorithm A.2 to represent any appendage shapes with octree modeling. Otherwise, calculate contribution to gravity field by shape $\mathbf{S}_i$ in convenient coordinate system $\{(\phi_{1i}(x),\phi_{2i}(y),\phi_{3i}(z))\}_i$ which allows for max of 1, 2, or 3 symbolic integrations in Equation \ref{eq:defpot} during potential $V$ calculation. In calculation of gravitational force $F$, take the derivative of Equation \ref{eq:defforce} evaluated at given shape $\mathbf{S}_i$, which is the potential $V$ contributed by $\mathbf{S}_i$, prior to integrating. Then integrate, in the choice coordinates $\{(\phi_{1i}(x),\phi_{2i}(y),\phi_{3i}(z))\}_i$. We encourage that these calculations to be migrated and done in parallel via GPU.
    \item \textbf{Compute Overall Gravity Field.}  Determine overall gravitational potential $V(\overrightarrow{\mathbf{x}})$ and gravitational force $F(\overrightarrow{\mathbf{x}})$ by again using the gravitational calculus to sum up all component pieces contributed by all shape pieces $\mathbf{B}$ and \{$\mathbf{S}_i$\}$_i$ from Step 3.
    \end{enumerate}
                    }%
    }

\begin{figure}[h!]
\centering
\includegraphics[scale = .5]{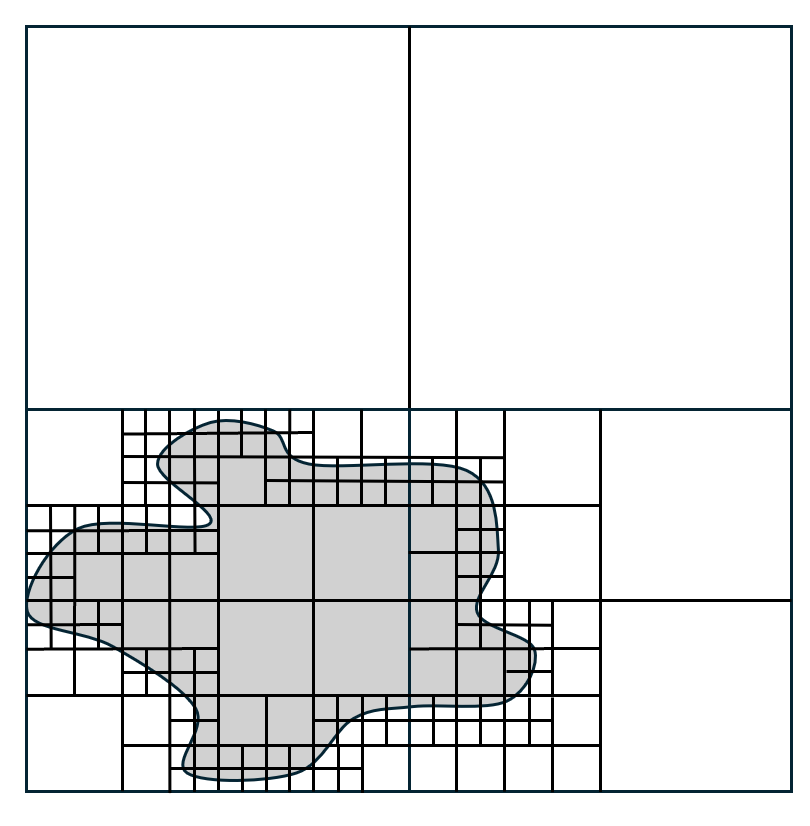}
 \caption{Shape $\mathbf{S}$ in gray represents 2-dimensional cross-section of shape of interest. Quad-tree algorithm out to 5 steps is demonstrated. Generalizes to 3 dimensions in octree representation. See Algorithm A.2 for the octree representation method.}
    \label{fig:qtree}
\end{figure}

\clearpage
\noindent\fbox{%
  \parbox{\dimexpr\linewidth-2\fboxsep-2\fboxrule\relax}{ \textbf{\large{Algorithm A.2: Gravitational Potential $V$, Gravitational}} \\ \textbf{\large{Force $F$ Computation using Octree Shape Representation}} \label{tab:alg2}
  \newline ||||||||||||||||||||||||||||||||||||
  \begin{enumerate}[leftmargin=*]
    \item   \textbf{Octree algorithm.} Begin with box that fully encapsulates the shape of interest $\mathbf{S}$. Then execute an octree algorithm out to chosen tolerance, whereby boxes are split into eight quadrants whenever there is mass inside a given box on a step of the algorithm. Stop at ``N steps" when tolerance is reached. Use a plug-in software for octree shape representation. See Figure \ref{fig:qtree} for 2-dimensional cross-section view of this process, which results in any shape representation \citep{nikolov2022}.
    \item  \textbf{Use of other octrees (if necessary).} One may choose to use pre-determined octree data structure computations for pieces of the shape. For instance, crater shapes may already have a pre-computed octree representation. See Algorithms A.3.1 and A.3.2 to move shape representation via calculus to desired location. 
    \item \textbf{Compute shape gravitational potential and force.} First compute closed-form gravitational potential and force of each box size used in the octree, at the origin. This corresponds to at most the number of steps in the algorithm (``N" closed-form box computations). See Equation \ref{eq:boxpot} for the closed-form representation of the potential induced at point $\overrightarrow{\mathbf{P}} = (P_1, P_2, P_3)$ by a box of dimensions 2$D_1$, 2$D_2$, 2$D_3$ in $x, y$, and $z$ respectively, centered at the origin, as given by \cite{chappell2012gravitationalfieldcube}. Equation \ref{eq:boxpot} uses the substitution $C = p_1p_2p_3$. The closed-form force field representation is given by the negative gradient of the potential, as given in Equation \ref{eq:boxforce}. Next operate by translation and rotation as in Algorithms A.3.1 and A.3.2, and sum up all components of boxes which contain mass of the body, to find the resulting gravitational potential and gravitational force field of the desired shape. The resulting gravity field is the sum of contributions from all the boxes with mass as given by Equation \ref{eq:boxsumv} and Equation \ref{eq:boxsumf}.
    \item \textbf{Representing densities within boxes.} In future work, non-constant shape densities may be inferred within each box using power series machine learning methods, but this requires further investigation.
    \end{enumerate}
                    }%
    }

\begin{equation} \label{eq:boxpot}
        V_B(\overrightarrow{\mathbf{P}}) = -G\rho \sum_{j=1}^{3} \begin{bmatrix} \ \sum_{i=1}^{3} 
        \begin{bmatrix}
        \frac{C}{p_i} \ln{(p_i+1)}\\
        -\frac{p_i^2}{2} \arctan{\frac{C}{p_i^2||\overrightarrow{\mathbf{p}}||}}
        \end{bmatrix} 
        \end{bmatrix}^{D_j-P_j}_{p_j=-D_j-P_j}
\end{equation}

\begin{equation} \label{eq:boxforce}
        F_B(\overrightarrow{\mathbf{P}}) = -\nabla 
        [V_{B}(\overrightarrow{\mathbf{P}})]
\end{equation}

\begin{equation} \label{eq:boxsumv}
        V(\overrightarrow{\mathbf{P}}) =  \sum_{B} 
        V_{B}(\overrightarrow{\mathbf{P}})
\end{equation}

\begin{equation} \label{eq:boxsumf}
        F(\overrightarrow{\mathbf{P}}) =  \sum_{B} 
        F_{B}(\overrightarrow{\mathbf{P}})
\end{equation}
        
\noindent\fbox{%
  \parbox{\dimexpr\linewidth-2\fboxsep-2\fboxrule\relax}{ \textbf{\large{Algorithm A.3.1: Gravitational Calculus; Translation}} \label{tab:gct}
  \newline ||||||||||||||||||||||||||||||||||||
  \begin{enumerate}[leftmargin=*]
    \item   \textbf{Original potential and force.} Take given gravitational potential $V(x,y,z) = V(\overrightarrow{\mathbf{{p}}})$ and force $F(x,y,z)  = F(\overrightarrow{\mathbf{{p}}})$ of shape $\mathbf{S}$ in the home coordinate system of $\mathbf{S}$, centered at the origin. Take desired translated shape $\mathbf{S_T}$. Determine vector $\overrightarrow{\mathbf{{p_0}}} = (x_{0},y_{0},z_{0})$ separating the center of the congruent shapes $\mathbf{S}$ and $\mathbf{S_T}$.
    \item  \textbf{Compute translated gravitational force and potential.} To compute translated gravitational  potential $V_{T}(\overrightarrow{\mathbf{{p}}})$, use $V_{T}(\overrightarrow{\mathbf{{p}}}) = V(\overrightarrow{\mathbf{{p}}}-\overrightarrow{\mathbf{{p_0}}}) = V(x-x_0,y-y_0,z-z_0)$. Analogously, to compute translated gravitational force $F_{T}(\overrightarrow{\mathbf{{p}}})$, use the operation $F_{T}(\overrightarrow{\mathbf{{p}}}) = V(\overrightarrow{\mathbf{{p}}}-\overrightarrow{\mathbf{{p_0}}}) = F(x-x_0,y-y_0,z-z_0)$.
    \end{enumerate}
                    }%
    }

\noindent\fbox{%
  \parbox{\dimexpr\linewidth-2\fboxsep-2\fboxrule\relax}{ \textbf{\large{Algorithm A.3.2: Gravitational Calculus; Rotation}}
  \newline ||||||||||||||||||||||||||||||||||||
  \begin{enumerate}[leftmargin=*]
    \item   \textbf{Original potential and force.}  Take given gravitational potential $V(x,y,z) = V(\overrightarrow{\mathbf{{p}}})$ and force $F(x,y,z)  = F(\overrightarrow{\mathbf{{p}}})$ of shape $\mathbf{S}$ in the home coordinate system of $\mathbf{S}$, centered at the origin. Take desired rotated shape $\mathbf{S_R}$. Determine rotation matrix \textbf{A}$\in$SO(3) such that for each $\overrightarrow{\mathbf{{p}}}\in\mathbf{S}$, the transformation $\textbf{A}\cdot\overrightarrow{\mathbf{{p}}}$  maps onto $\mathbf{S_R}$.
    \item   \textbf{Compute rotated gravitational force and potential.} To compute rotated gravitational potential $V_{R}(\overrightarrow{\mathbf{{p}}})$, use $V_{R}(\overrightarrow{\mathbf{{p}}}) = V(\textbf{A}^{-1}\cdot\overrightarrow{\mathbf{{p}}}) = V(\textbf{A}^{-1}\cdot(x,y,z))$. Analogously, to compute translated gravitational force $F_{R}(\overrightarrow{\mathbf{{p}}})$, use the operation $F_{R}(\overrightarrow{\mathbf{{p}}}) = F(\textbf{A}^{-1}\cdot\overrightarrow{\mathbf{{p}}}) = F(\textbf{A}^{-1}\cdot(x,y,z))$.
    \end{enumerate}
                    }%
    }

\begin{figure}[h!]
\centering
\includegraphics[scale = .5]{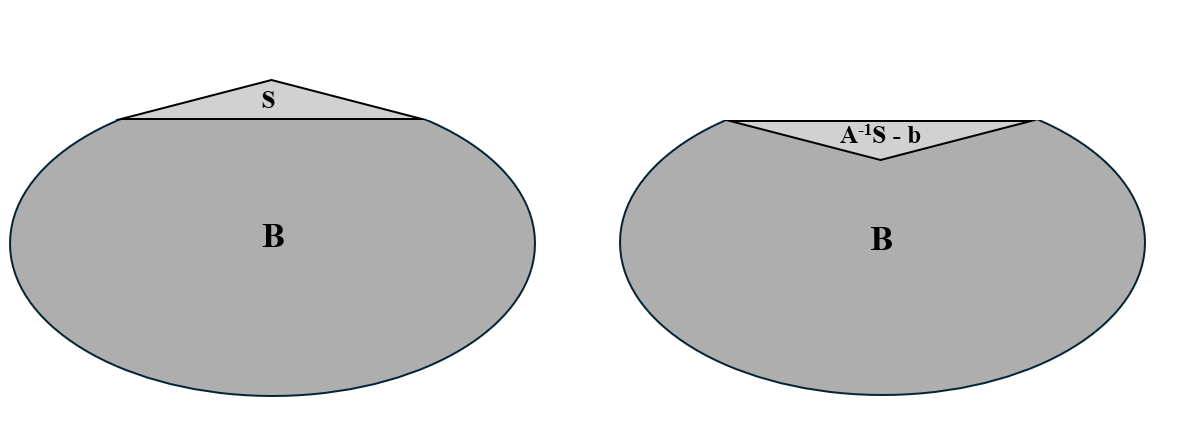}
 \caption{Consider these shapes as cross-sections of a 3-dimensional ellipsoid with appendage shape $\mathbf{S}$, where shape $\mathbf{S}$ represents a flat cone which is intersected with base shape ellipsoid $\mathbf{B}$. The intersection region $\mathbf{S_I}$ is the section of the ellipsoid that is missing, and this region is ``subtracted" using the gravitational calculus to find the gravitational potential and force fields of the right and left object. The action of a rotation and translation on shape $\mathbf{S}$ returns a crater without needing to recompute $\mathbf{S}$, using the gravitational calculus. For the left shape, the gravity field is the result of the gravity field of $\mathbf{B}$ minus the gravity field of $\mathbf{S_I}$, the piece of $\mathbf{B}$ which is chopped off, \emph{plus} the gravity field of $\mathbf{S}$. The right shape's gravity field is the gravity field of $\mathbf{B}$ minus the gravity field of $\mathbf{S_I}$, \emph{minus} the gravity field of $\mathbf{S}$.}
    \label{fig:flatcone}
\end{figure}

\noindent\fbox{%
  \parbox{\dimexpr\linewidth-2\fboxsep-2\fboxrule\relax}{ \textbf{\large{Algorithm A.4: Gravitational Calculus; Re-Use of Volumes}}
  \newline ||||||||||||||||||||||||||||||||||||
  \begin{enumerate}[leftmargin=*]
    \item   \textbf{Given ``convenient" shapes to use.} Use shapes $\mathbf{S}$, $\mathbf{S_I}$ such that removing the region $\mathbf{S_I}$ from any number of places in the given base shape $\mathbf{B}$, which may be a sphere or ellipsoid, allows for re-use of the shape $\mathbf{S}$ with rotation and translation to model the features of the shape of interest. 
    \item   \textbf{Move the shape pieces to the desired locations.} Move $\mathbf{S}$ to the (multiple) locations of computation via translation and/or rotation and compute the gravity field of the base shape piece (or pieces) that has been removed, $\mathbf{S_I}$.
    \item   \textbf{Compute overall gravity field.} Compute the gravitational potential and force of the shape of interest which is formed with just combinations of $\mathbf{S_I}$, $\mathbf{S}$, and $\mathbf{B}$ being careful not to double count or double subtract. See Figure \ref{fig:flatcone}. Refer to Algorithms 2-3 for computation in $\mathbf{S}$, $\mathbf{S_I}$.
    \end{enumerate}
                    }%
    }

\clearpage

\begin{figure}[h!]
\centering
\includegraphics[scale = .5]{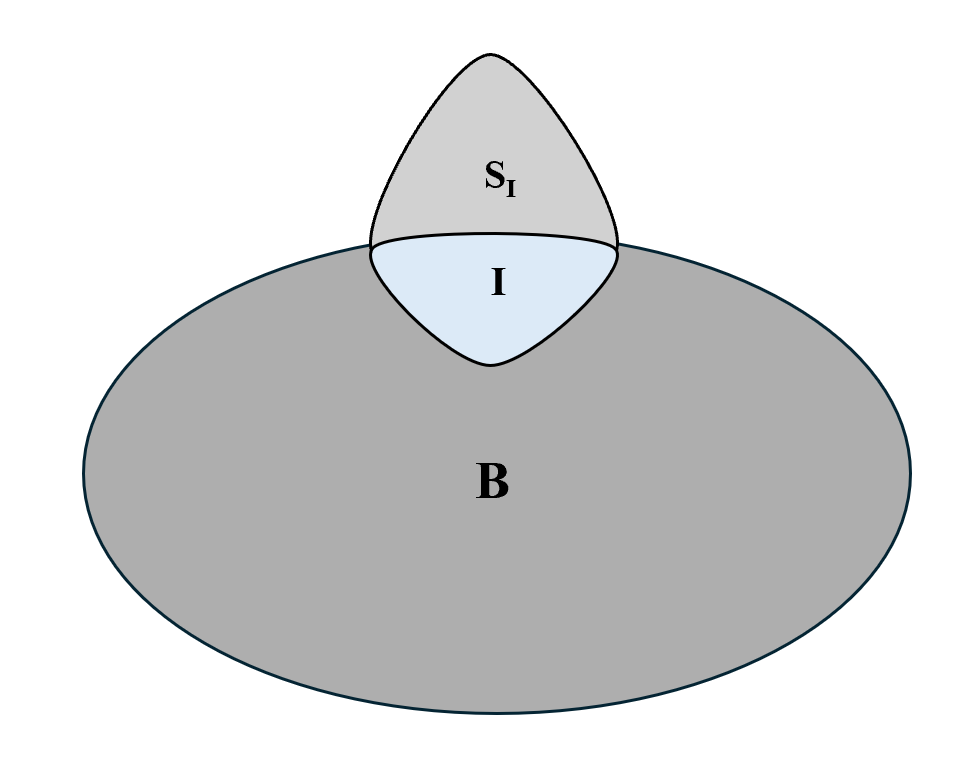}
 \caption{Shape $\mathbf{S}_I$ in gray and blue is intersected with base shape $\mathbf{B}$. Intersection region is $\mathbf{I}$. The intersection region resembles that of a crater. }
    \label{fig:intr}
\end{figure}

\noindent\fbox{%
  \parbox{\dimexpr\linewidth-2\fboxsep-2\fboxrule\relax}{ \textbf{\large{Algorithm A.5: Gravitational Calculus; Intersection}}
  \newline ||||||||||||||||||||||||||||||||||||
  \begin{enumerate}[leftmargin=*]
    \item   \textbf{Determine a ``convenient" shape to use.} Determine shape $\mathbf{S}_I$ such that intersection derives a wide combination of shapes which represent the surface features of the desired shape $\mathbf{S}$, upon intersection with a chosen base shape $\mathbf{B}$, which may be a sphere or ellipsoid.
    \item   \textbf{Move the shape to the desired location.} Move $ \mathbf{S}_I$ to location of computation via translation, rotation, and/or scaling.
    \item   \textbf{Compute at intersection.} Compute the gravitational potential and force of the intersection $\mathbf{I}$ between $ \mathbf{S}_I$ and the base shape $\mathbf{B}$; see Figure \ref{fig:intr}. Refer to Algorithms 2-3 for computation in $\mathbf{I}$. 
    \end{enumerate}
                    }%
    }

\textbf{Packages: GravToolbox. } Presented in MATLAB, this directory gives a set of files which can compute specified crater or base sphere gravity fields in cylindrical coordinates. The package also generalizes as a convenient tool for computations of shapes which are a linear combination of craters and a base shape. 
\textbf{Potential Reconstruction Algorithm.} Mathematica Demonstration of Algorithm 1 versus direct integration. The package features a presentation of a current state-of-the-art implementation on a sphere, where there is some error as shown in Figure \ref{fig:error}, while the numerical method used has extremely low error.

\section{Appendix B} \label{sec:ap2}
Appendix B analyzes the challenges that arise in conventional methods of gravitational potential field (here denoted by $U$ to match previous notations) and gravitational force field ($\nabla U$) calculations. We also present an implementable algorithm (B.1) giving the state-of-the-art fast computation method. Most methods use spherical harmonics expansions, though ellipsoidal expansions have also been used, and in certain cases improved accuracy \citep{2001ellipsoid}. These, however, are each derived in such a way that require assumptions in the posing of the problem. 

\textbf{Conventional Aerospace Method Challenges: } Conventional state-of-the-art aerospace methods use series expansions of solutions to the Laplace Equation, given by Equation \ref{eq:laplace}, which is solved by separation of variables to determine orthonormal basis functions which determine the spherical harmonics basis functions whose coefficients are conventionally given by Equation \ref{eq:ecoeff} \citep{2014Takahashi}, \citep{scheeres2016orbital}, \citep{2015ZB}. This gives approximations for the gravitational potential $U^e$ outside the body given by Equation \ref{eq:outpot}.
\begin{equation} \label{eq:laplace}
\nabla^2U = 0
\end{equation}

\begin{equation} \label{eq:ecoeff}
\left\{
\begin{aligned}
    C^e_{nm} &= \frac{(2 - \delta_{0m})(n - m)!}{M * (n + m)!} \int_M \left( \frac{r'}{R_e^*} \right)^n P_{nm}(\sin\phi') \cos(m\lambda') \, dm' \\
    S^e_{nm} \big|_{m > 0} &= \frac{2(n - m)!}{M * (n + m)!} \int_M \left( \frac{ r'}{R_e^*} \right)^n P_{nm}(\sin\phi') \sin(m\lambda') \, dm'
\end{aligned}
\right.
\end{equation}

\begin{equation} \label{eq:outpot}
U^e(r, \lambda, \phi) = \frac{GM^*}{R_e^*} \sum_{n=0}^{\infty} \sum_{m=0}^{n} P_{nm} (\sin \phi) 
\begin{bmatrix}
\cos(m\lambda) \\
\sin(m\lambda)
\end{bmatrix} \cdot 
\begin{bmatrix}
C^e_{nm} \\
S^e_{nm}
\end{bmatrix}
\end{equation}

This solution method works best for the homogeneous differential equation (Equation \ref{eq:laplace}) and is effective for solving for the gravity field of the body. To calculate the potential or force at points interior to the shape of interest, Poisson's Equation is conventionally solved, as given by Equation \ref{eq:poisson}. 

\begin{equation} \label{eq:poisson}
\nabla^2U = -4\pi G\rho
\end{equation}

Basis functions for the interior case have conventionally been found as Bessel function coefficients presented by Equation \ref{eq:icoeff}, though this work assumes that gravitational potential $V$ is proportional to the density distribution \citep{2014Takahashi}.
\begin{equation} \label{eq:icoeff}
\left\{
\begin{aligned}
    \mathscr{A}_{lnm}^i &= \frac{2(2 - \delta_{0m})(2n + 1)(n - m)!}{M * \mathscr{E}_n^i (\alpha_{ln}^i) (n + m)!} \int_M j_n \left( \frac{\alpha_{ln}^i r'}{R_e^*} \right) P_{nm}(\sin\phi') \cos(m\lambda') \, dm' \\
    \mathscr{B}_{lnm}^i \big|_{m > 0} &= \frac{4(2n + 1)(n - m)!}{M * \mathscr{E}_n^i (\alpha_{ln}^i) (n + m)!} \int_M j_n \left( \frac{\alpha_{ln}^i r'}{R_e^*} \right) P_{nm}(\sin\phi') \sin(m\lambda') \, dm'
\end{aligned}
\right.
\end{equation}

\begin{equation} \label{eq:inpot}
U^i(r, \lambda, \phi) = \frac{GM^*}{R_e^*} \sum_{l=0}^{\infty} \sum_{n=0}^{\infty} \sum_{m=0}^{n} j_n \left( \frac{\alpha_{ln}^i r}{R_e^*} \right) P_{nm} (\sin \phi) 
\begin{bmatrix}
\cos(m\lambda) \\
\sin(m\lambda)
\end{bmatrix} \cdot 
\begin{bmatrix}
\mathscr{A}_{lnm}^i \\
\mathscr{B}_{lnm}^i
\end{bmatrix}
\end{equation}

This state-of-the-art procedure for both the interior and exterior cases in this procedure are summarized by Algorithm B.1.

\begin{figure}
\centering
\includegraphics[scale = .725]{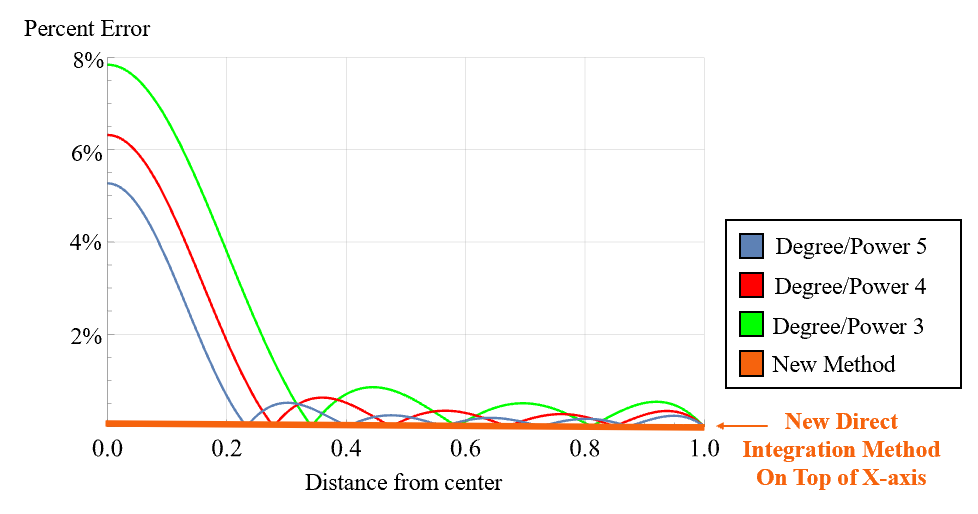}
 \caption{Error in computational methods on the uniform density sphere. Our method of solving the integral directly only has numerical error, which is bounded by $10^{-11}$ percent, indistinguishable from the x-axis. ``Degree/Power m" expansions are state-of-the art spherical harmonic expansions of the form given in \citep{2014Takahashi} where the expansion is up to degree ``m" and power ``m." See Equation (\ref{eq:inpot}), where the first and second infinite summations can be truncated to a finite bound which correspond to the power and degree, respectively.}
    \label{fig:error}
\end{figure}

\textbf{Main Sources of Error: }There is a key region where the errors in current state-of-the-art methods arise in the calculation of the gravitational force field for any non-spherical body. Namely, assuming we have some mass $M$ with non-negligible volume $V$, a Brillouin sphere is any sphere centered at the center of $M$ which entirely encloses $M$. Within the Brillouin sphere, conventional gravitational force predictions are generally demonstrated to be inaccurate and divergent \citep{costin2020}, \citep{SPRLAK2021103739}. Even for ``nearly spherical" bodies, the divergence of the predication proves to be significant; historically, state-of-the-art methods give errors on the order of tens of percent for calculations within the
minimum Brillouin sphere of the body \citep{2014Takahashi}, \citep{scheeres2016orbital}, \citep{SCHEERES2019}. See Figure \ref{fig:error} for a comparison of these methods with our method of direct integration, about a uniform mass sphere. 

Spherical harmonics expansions are a standard approximation to the gravitational field far from the body, but, in fact, convergence of spherical harmonics occurs with probability zero inside the minimum Brillouin sphere \citep{costin2020}. Even for bodies such as the moon, where spherical harmonics expansions and other smooth approximations have been used throughout literature, direct surface measurements taken by GRAIL (Gravity Recovery and Interior Laboratory) indicate divergences and significant inaccuracy to conventional aerospace calculations within the minimum Brillouin sphere \citep{SPRLAK2021103739}. Our problem lies in that the ``cusp-like" characteristics of the true force function make smooth approximations highly inaccurate near the surface; that is, the true force function is riddled with cusps and non-differentiable points as was shown in Figure \ref{fig:sgrav} for the gravitational force field near a sphere. 
\noindent\fbox{%
  \parbox{\dimexpr\linewidth-2\fboxsep-2\fboxrule\relax}{ \textbf{\large{Algorithm B.1: Gravitational Potential $U(r, \lambda, \phi)$}} \\ \textbf{\large{Computation (State-of-the-art, \cite{2014Takahashi})}}\label{tab:alg1}
  \newline ||||||||||||||||||||||||||||||||||||
  \begin{enumerate}[leftmargin=*]
    \item  \textbf{Eigenvalues $\alpha^i_{ln}$.} Compute interior (Bessel) gravity field eigenvalues $\alpha^i_{ln}$ via boundary condition $j_{n-1}(\alpha^i_{ln}) = 0$ upon defining $j_n$ as the spherical Bessel function of the first kind, letting $n$ be the degree and $l$ be the power. Refer to Table \ref{tab:eigdata} for pre-computed values.
    \item  \textbf{Normalization factors $\mathscr{E}^i_{q}$, constants $G$, $M$, $R^*_e$.} Working in spherical coordinates, define $\mathscr{E}_q^i (\alpha_{ln}^i) = \alpha_{ln}^{i2} \left[ j_q(\alpha_{ln}^i) \right]^2$, latitude $\lambda$, longitude $\phi$, volume element $dv' = r'^2 \sin\theta \, dr' \, d\phi' \, d\lambda'$ via integration over points $(r', \lambda', \phi')$ with nonzero density, and colatitude $\theta = \pi/2 - \phi$. Define constants $M$ to be reference mass of the body, $R^*_e$ the reference radius of the body, and $G$ the gravitational constant. Define $P_{nm}$ to be the Legendre polynomials of order n and degree m. 
    \item \textbf{Interior gravity field basis coefficients $\mathscr{A}^i_{lnm}$, $\mathscr{B}^i_{lnm}$.} Compute interior (Bessel) gravity field non-dimensional coefficients $\mathscr{A}^i_{lnm}$, $\mathscr{B}^i_{lnm}$ via Equation (\ref{eq:icoeff}) and scale them by the normalization factor $R^*/M^*G$. 
    \item \textbf{Exterior gravity field basis coefficients $C^e_{nm}$, $S^e_{nm}$.} Compute exterior gravity field non-dimensional coefficients $C^e_{nm}$, $S^e_{nm}$ via Equation (\ref{eq:ecoeff}).
    \item \textbf{Interior gravity field potential.} Define interior gravity potential field $U^i$ as in Equation (\ref{eq:inpot}); using previous definitions and computations, this provides an analytical expression for $U^i$ as a function of any point $(r, \lambda, \phi) \epsilon \mathbf{R}^3$ up to $r<R^*_e$.
    \item \textbf{Exterior gravity field potential.} Define exterior gravity potential field $U^e$ as in Equation (\ref{eq:outpot}) for an analytical expression of potential at any point $(r, \lambda, \phi) \epsilon \mathbf{R}^3$ such that $r \geq R^*_e$.
    \end{enumerate}
                    }%
    }
\begin{table}[h!]
    \centering
    \caption{Interior Bessel Gravity Field Eigenvalues $\alpha^i_{ln}$. See \cite{2014Takahashi} for computation.}
    \label{tab:eigdata}
    \begin{tabular}{|c|c|c|c|c|c|c|}
        \hline
        \textbf{(Power l, Deg. n)} & \textbf{Deg. 0:} & \textbf{Deg. 1:} & \textbf{Deg. 2:} & \textbf{Deg. 3:} & \textbf{Deg. 4:} & \textbf{Deg. 5:} \\
        \hline
        \textbf{Power 0:} & 1.5708 & 3.1416 & 0 & 0 & 0 & 0 \\
        \hline
        \textbf{Power 1:} & 4.7124 & 6.2832 & 4.4934 & 5.7635 & 6.9879 & 8.1826 \\
        \hline
        \textbf{Power 2:} & 7.8540 & 9.4248 & 7.7253 & 9.0950 & 10.4171 & 11.7049 \\
        \hline
        \textbf{Power 3:} & 10.9956 & 12.5664 & 10.9041 & 12.3229 & 13.6980 & 15.0397 \\
        \hline
        \textbf{Power 4:} & 14.1372 & 15.7080 & 14.0662 & 15.5146 & 16.9236 & 18.3013 \\
        \hline
        \textbf{Power 5:} & 17.2788 & 18.8496 & 17.2208 & 18.6890 & 20.1218 & 21.5254 \\
        \hline
    \end{tabular}
\end{table}

\clearpage
\bibliography{refs.bib}{}
\bibliographystyle{aasjournal}
\end{document}